## RESEARCH ARTICLE

# The First Comparison Between Swarm-C Accelerometer-Derived Thermospheric Densities and Physical and Empirical Model Estimates

Timothy Kodikara[1,2], Brett Carter[1], and Kefei Zhang[1]

[1]SPACE Research Centre, RMIT University, Melbourne, Victoria, Australia, [2]SERC Limited, AITC2 Mount Stromlo Observatory, Canberra, ACT, Australia

**Abstract** The first systematic comparison between Swarm-C accelerometer-derived thermospheric density and both empirical and physics-based model results using multiple model performance metrics is presented. This comparison is performed at the satellite's high temporal 10-s resolution, which provides a meaningful evaluation of the models' fidelity for orbit prediction and other space weather forecasting applications. The comparison against the physical model is influenced by the specification of the lower atmospheric forcing, the high-latitude ionospheric plasma convection, and solar activity. Some insights into the model response to thermosphere-driving mechanisms are obtained through a machine learning exercise. The results of this analysis show that the short-timescale variations observed by Swarm-C during periods of high solar and geomagnetic activity were better captured by the physics-based model than the empirical models. It is concluded that Swarm-C data agree well with the climatologies inherent within the models and are, therefore, a useful data set for further model validation and scientific research.

**Plain Language Summary** The first systematic comparison of the thermospheric densities between Swarm-C accelerometer-derived densities is presented. The data are compared to two latest versions of a physics-based general circulation model and arguably the most widely used empirical models using multiple model performance metrics. The comparison at the satellite's temporal resolution provides a useful evaluation of the models' fidelity for orbit prediction and pertinent space weather forecasting applications. The results of this study show that the short-timescale variations observed by Swarm-C during periods of high solar and geomagnetic activities were better captured by the physics-based model than the empirical model. The results show the complex interconnectedness of solar activity and geomagnetic activity on model performance in terms of estimating density. The study also uses a machine learning exercise to demonstrate characteristics inherent to each model run. Swarm-C data are consistent with the models. It is concluded that Swarm-C data are suitable for further model validation and scientific research.

## 1. Introduction

The response of neutral mass density in the thermosphere (hereinafter density) to various space weather conditions is not fully understood. Density is one of the most significant uncertainties in tracking and predicting the orbit of a space object in low Earth orbit (LEO; Bennett et al., 2015; Vallado, 2004; Vallado & Finkleman, 2014). Precise knowledge of the variations in density is also critical for attitude control, predicting satellite lifetime and reentry point as well as collision avoidance maneuvers in LEO (Vallado, 2004). The only way to predict density in space and time is through modeling, ergo the uncertainty in density estimates translates into safety margins applied for satellites mainly in LEO (J. Chen et al., 2017; Zesta & Huang, 2016). Emmert et al. (2017) provide several examples of stochastic uncertainty in density being propagated to predictions of satellite orbits.

Various techniques to estimate density exist and have been appraised in many studies (e.g., Emmert, 2015). Empirical models of density are based on climatologies derived from observations. Although most widely used in orbit tracking and prediction exercises, empirical models are limited in their resolution especially at short timescales mainly due to the averaging and scarcity of observations (Emmert, 2015). Physical models numerically solve the fluid equations to derive evolving thermospheric parameters such as density, temperature, winds, and composition (Qian & Solomon, 2012). Capturing the evolving thermospheric responses is critical for a forecast model. For example, as noted in Lin et al. (2013), satellite tracking errors at an orbit altitude of 400 km during even moderate magnetic storms are about 65% greater than at quiet times. Looking at the growth and resilience regarding nowcast and forecast capabilities in physical models of terrestrial weather, atmospheric researchers are sanguine about similar prospects for physics-based upper atmospheric models (Y. Chen et al., 2011).

Two main drivers of density variations are the solar irradiance and the geomagnetic activity (Knipp et al., 2004). In modeling the contribution of these drivers, both physics-based Thermosphere-Ionosphere-Electrodynamics General Circulation Model (TIE-GCM; see section 2.1) and the empirical Mass Spectrometer Incoherent Scatter Radar Model (NRLMSISE-00; see section 2.2) use $F_{10.7}$ (see Tapping, 2013) as a proxy for the solar irradiance and $Kp$ and $Ap$ indices (see Menvielle & Berthelier, 1991), respectively, to represent the geomagnetic activity. Also, internal mechanisms originating from the lower atmosphere, such as tides, turbulence, and planetary waves, dissipate energy into the ionosphere-thermosphere system and thus change its thermal, dynamical, and compositional structure (Liu, 2016). At the TIE-GCM's lower boundary an empirical formulation representing the seasonal variations in the advective and diffusive transport of primarily atomic oxygen is imposed (Qian et al., 2009). While this representation of lower atmospheric disturbances (collectively referred to as eddy diffusion [$K_{zz}$]) showed an improvement in Qian et al. (2009), it also engendered undesired problems in Siskind et al. (2014) and Jones et al. (2017) as $K_{zz}$ attempts to contain both short wavelength eddies and larger dynamical and diffusive processes.

Codrescu et al. (2008) compared the impact of $F_{10.7}$ and $Kp$ on thermospheric temperature using the physics-based coupled thermosphere ionosphere plasmasphere electrodynamics (CTIPE) model (see Fuller-Rowell & Rees, 1980) and the MSIS-86 model, which is an earlier version of the NRLMSISE-00, and concluded that irrespective of the season and geomagnetic activity, on average the CTIPE model underestimates temperature at $F_{10.7}$ flux values ranging from low to high compared to the MSIS-86 model. Although the time-of-day dependency on the model performance was small in Codrescu et al. (2008), the temperature variation due to $F_{10.7}$ in the CTIPE model was influenced by both season and geomagnetic activity. Navier-Stokes equations applied to the thermosphere relate that density depends on the temperature profile as well as the compositional structure of ions and neutrals. Moreover, Masutti et al. (2016) showed that the density bias in the physics-based global ionosphere-thermosphere model (see Ridley et al., 2006) compared to CHAMP (Challenging Minisatellite Payload) and GRACE-A (Gravity Recovery and Climate Experiment) accelerometer-derived densities linearly increases with $F_{10.7}$. Such a linearly increasing bias with $F_{10.7}$ has not been reported for TIE-GCM. Emmert et al. (2014) provide a comparison of average density change during two consecutive solar minima due to $F_{10.7}$ and $Kp$ in NRLMSISE-00 and TIE-GCM version 1.94.2 with respect to a statistical formulation in the global average mass density model. Emmert et al.'s (2014) study concluded that the effect attributable to $F_{10.7}$ and $Kp$ is not significantly different for both models, NRLMSISE-00 and TIE-GCM, under quiet conditions. Rather than using the $F_{10.7}$, Solomon et al. (2011), using the MgII core-to-wing ratio as the solar extreme ultraviolet (EUV) proxy to TIE-GCM version 1.93, showed that EUV effect on density change during the 2007–2009 solar minimum is significantly higher than that of the geomagnetic activity.

Siemes et al. (2016) showed the good correspondence between calibrated Swarm-C density and solar and geomagnetic activities indicating the suitability of the data for detailed investigations. Swarm-C satellite (see Friis-Christensen et al., 2008) is equipped with eight channels of Global Positioning System (GPS) trackers. The GPS signals combined with precise orbit determination techniques are used to calibrate and validate the accelerometer data. Xiong et al. (2016) report on the total loss of GPS signal events during 2013–2015 largely encountered near the equator at local nighttimes (19:00–22:00). Such losses for Swarm-C have not been reported since May 2015 following an update to receiver settings (Xiong et al., 2016). The effect of these total loss of GPS signal events on the accelerometer-derived density is yet to be determined. The task of extracting the pure nongravitational signal from the raw accelerometer measurements is discussed in Siemes et al. (2016). Bezděk et al. (2017) provide a comparison between these raw measurements and modeled nongravitational accelerations based on NRLMSISE-00-derived densities. Yet, to our knowledge, accelerometer-derived densities from the Swarm-C have not been used, so far, as a viable model validation data set.

In most of these model performance analyses, the model estimates and data are normalized to a nominal altitude and averaged out hourly or daily (e.g., Elvidge et al., 2016; Emmert et al., 2014). The biases induced by the normalization technique are canceled out when both data and model estimates are treated equally. Therefore, normalization simplifies the comparison and is rather suitable for a comparison of the average behavior of model and data. As noted in Elvidge et al. (2014), assessing model performance using just one metric (typically difference or ratio between model and data) may hide other underlying biases of the model. Furthermore, when comparing long-term density with, for example, solar or geomagnetic activities, data/model ratio provides a better visualization of the correlation than the residuals as the amplitude difference of density (over seasons, years, etc.) are higher than the amplitude difference of the solar and geomagnetic activities. In other words, model-data density differences during higher-density periods may eclipse the small differences during lower-density periods. A comparison along each epoch, on the other hand, may provide a more robust test of the ability of a forecast model to describe the high-resolution temporal and spatial variations.

In this study, the model performance is analyzed at the Swarm-C's spatial and temporal resolution by using five different metrics. This paper provides a useful validation of the new Swarm-C data and shows the valuable addition it makes to the existing density database, which is sparse compared to other regions of the atmosphere. The paper also discusses the effect of driving NRLMSISE-00 with the indicated full history of $Ap$ and the daily average $Ap$ under solar maximum conditions. The impact of space weather parameterization ($F_{10.7}$ and $Kp$) in TIE-GCM is investigated using various techniques. The impact of $K_{zz}$ on density is studied using the latest version of TIE-GCM along with a comparison of two high-latitude empirical ion convection models. Furthermore, utilizing a machine learning cross-validation scheme, how the systematic bias due to space weather parameterization manifests in models is also studied.

## 2. Models and Data

The study presented here covers approximately 1 year (June 2014 to May 2015) of the recent solar maximum period with the lowest recorded solar activity in over 100 years since record keeping began circa 1750 (Hathaway, 2015). A brief description of the Swarm-C data and the models used in the comparison is provided below.

The TIE-GCM was chosen to compare the data against a well-established self-consistent physics-based model (e.g., Emmert, 2015; Qian et al., 2014) to obtain an understanding of the data with respect to internal physics of the thermosphere system. The two empirical models, NRLMSISE-00 and DTM-2013 (Drag Temperature Model), were chosen to represent a sample of the most comprehensive empirical density models (see Emmert, 2015) that are also widely used in applications of orbit determination and prediction (e.g., Bruinsma, 2015; Emmert et al., 2017; McLaughlin et al., 2011).

### 2.1. Thermosphere-Ionosphere-Electrodynamics General Circulation Model

TIE-GCM is a three-dimensional, time-dependent, physical model of the upper atmosphere (Dickinson et al., 1981; Richmond et al., 1992). TIE-GCM assumes a constant acceleration due to gravity (870 cm/s$^2$) and spherical symmetry of the Earth. This study employs the two most recent versions of the TIE-GCM: 1.95 and 2.0. The latest version distinguishes argon as a minor constituent and helium as a major constituent. The helium concentration can have a significant impact on atmospheric drag calculations, especially at high latitudes in the winter hemisphere (see Sutton et al., 2015, and references therein). Version 2.0 also allows the option to specify a background zonal mean climatology of winds and temperature at the lower boundary default flat conditions are used in our model runs.

The model runs are performed on a 5° × 5° grid in latitude and longitude along 29 isobaric layers that extend from approximately 97 to 600 km in altitude. The isobaric layers are separated by 0.5 $H$, where $H$ is the scale height of the constituents. Lower boundary wave forcing is specified through numerically derived migrating diurnal and semidiurnal tides (see Hagan et al., 1999).

TIE-GCM is capable of accepting magnetospheric inputs from multiple sources and methods, such as assimilating direct observations and coupling with empirical or numerical models. The inputs required are the high-latitude ion convection, hemisphere power, and the cross polar cap potential (CPCP). High-latitude ion convection patterns derived from the electric potential model of Weimer (2005, hereinafter TIE-GCM(W)) and the Heelis et al. (1982) ionosphere convection model (hereinafter TIE-GCM(H)) are utilized separately in the model runs presented here. High-latitude auroral precipitation is determined through $Kp$-dependent hemisphere power and CPCP. The Heelis et al. (1982) empirical model depends on the CPCP and the strength and direction of the interplanetary magnetic field (IMF). In addition to IMF strength and direction, the Weimer (2005) model depends on the dynamics of the solar wind and the orientation of the geomagnetic dipole.

The thermosphere-ionosphere coupling with the plasmasphere is described by the direction of flow of plasma flux at the upper boundary. The EUVAC (EUV flux model for aeronomic calculations) solar proxy model of Richards et al. (1994), with modifications by Solomon and Qian (2005), is used as the solar heat input throughout all the TIE-GCM simulations presented here. In the EUVAC model, solar heating is described by the $P$ index, which is a solar flux average based on the daily $F_{10.7}$ and its running 81-day centered mean that can be mathematically expressed as

$$P(i) = \frac{1}{2N+1} \sum_{j=-N}^{N} F_{10.7}(i+j), \tag{1}$$

where $i$ is the day of year that the model is evaluated for and $N$ is 40.

Each TIE-GCM simulation was primed with an arbitrarily chosen 15-day "settle-in" period, and only the model outputs after this period are considered here.

### 2.2. Mass Spectrometer Incoherent Scatter Radar Model

Picone et al. (2002) introduced the NRLMSISE-00 empirical model as a significantly modified version of the original MSIS-class models: MSIS-86 and MSIS-90 (see Hedin, 1987; Hedin et al., 1991). The NRLMSISE-00 model describes the atmosphere from ground level to the exosphere and is based on data from satellites, rockets, and radars over several decades (1961–1998; Picone et al., 2002). The model uses curve fitting techniques to estimate the temperature, composition, and density for a given altitude, latitude, longitude, universal time, 81-day centered mean $F_{10.7}$, daily $F_{10.7}$, $Ap$ geomagnetic indices for seven periods of anterior magnetic activity, and apparent local solar time.

One significant addition in the NRLMSISE-00 is the accounting of anomalous oxygen ($O^+$) in the calculation of density for altitudes beyond 500 km. Similar to most other upper atmospheric models, including TIE-GCM, the NRLMSISE-00 model accounts for the spherical symmetry of the Earth; however, unlike TIE-GCM, it uses spherical harmonics to resolve for the geographic coordinates to map the model outputs (Picone et al., 2002). In the NRLMSISE-00 model, the thermospheric outputs mainly depend on exospheric temperature profile for which the Walker (1965) temperature profile is used with certain modifications to the EUV contribution introduced by Picone et al. (2002).

### 2.3. DTM-2013

DTM-2013 has some major differences compared to the NRLMSISE-00 model. DTM-2013's reference database spans from 1961 to 2012, and the model approximately covers the altitudinal region of 120–1,500 km (Bruinsma, 2015). Unlike in NRLMSISE-00, in this study the $F_{30}$ (30-cm solar radio flux) proxy was used to describe the variation in solar activity for DTM-2013. Relevant to the context presented in this paper, Dudok de Wit et al. (2014) describe the significant differences between the $F_{30}$ and $F_{10.7}$ proxies. In DTM-2013, the geomagnetic variation is described as a function of 3-hr *am* index (see; Menvielle & Berthelier, 1991, and references therein) but also allows the use of $Kp$ as a substitute (Bruinsma, 2015).

The model takes into account the local time and latitude variation of density and temperature but not the variation across longitudes as a function of solar heating modulated by the $F_{30}$ solar proxy (Emmert, 2015). On top of density at a given time and location, DTM-2013 also provides the model uncertainty of the estimated value due to the low model resolution.

### 2.4. Swarm-C Accelerometer-Derived Neutral Density

Swarm, launched in late 2013, consists of three near-polar (angle of inclination: 87.4° [A and C]; 87.8° [B]) satellites designed for monitoring the magnetic field of the Earth. The satellites also carry accelerometers and GPS receivers on board measurements used to derive the density (Siemes et al., 2016).

The data product used in this analysis is the Swarm-C (average orbital height, 480 km) accelerometer-derived *Level2daily (DNSCWND)* postprocessed density product. The Swarm-C density product is subject to several disturbances and is not yet fully optimized to account for all variants of nongravitational accelerations (Siemes et al., 2016). In the Swarm constellation, only Swarm-C has been identified as the least affected by these disturbances. By March 2018, accelerometer-derived density from only the Swarm-C satellite was available. Only the linear acceleration (in-track) has been considered in the density product used in this study. It is expected that as the Swarm satellites gradually decay in orbital height, the higher nongravitational acceleration signal may help improve the calibration of accelerometer-derived densities.

## 3. Methods

The physics-based TIE-GCM and the empirical NRLMSISE-00, and DTM-2013 models were used in different configurations to estimate the densities along the Swarm-C orbit. The ensemble of model runs is listed below.

TIE-GCM runs:

1. T1 TIE-GCM(H)-version 2.0 with day-of-year-dependent $K_{zz}$ coefficient.
2. T2 TIE-GCM(H)-version 2.0 with constant $K_{zz}$ coefficient.
3. T3 TIE-GCM(W)-version 2.0 with constant $K_{zz}$ coefficient.
4. V1 TIE-GCM(H)-version 1.95 with constant $K_{zz}$ coefficient.

NRLMSISE-00 runs:

5. M1 NRLMSISE-00 driven with the full required history of $Ap$.
6. M2 Similar to M1 but with a daily average of 3-hr $Ap$.

DTM-2013 run:

7. D1 DTM-2013 driven with the default solar and geomagnetic indices.

Through Runs 1–4, the effect of different configurations of empirical high-latitude potential models coupled to the TIE-GCM, and $K_{zz}$ coefficient on density is investigated. Through Runs 5 and 6, the effect of accounting for empirically estimated $O^+$ and geomagnetic activity expressed as daily average or a combination of anterior magnetic activity up to 56 hr on density is studied using the NRLMSISE-00 model. The Run 7 compares the performance of DTM-2013 model against the rest.

The cubic spline interpolation scheme was used to map the TIE-GCM estimated densities along the satellite orbit, first along the latitude and longitude and then vertically to the satellite altitude. Normalizing the Swarm-C densities to a specific altitude was omitted to rid the data from interpolation biases and gain an opportunity to compare the models' performance at the satellite altitude, which is an important attribute in applications of orbit determination and orbit prediction. NRLMSISE-00 and DTM-2013 densities were estimated by passing the satellite coordinates to the model for each epoch serially.

Model bias in upper atmospheric density modeling is often assessed by different metrics, such as root-mean-square error (RMSE), standard deviation, absolute difference, model/observation ratio, log-normal ratio, percent change, or a combination thereof (e.g., Doornbos, 2012; Elvidge et al., 2014). The observed/model upper atmospheric density assumes a log-normal distribution and also exhibits approximate characteristics of a Gaussian (normal) distribution (Bezděk, 2007). However, the nonlinear error sources in data/model alter the log-normal mean of the distributions. Therefore, using the Gaussian standard deviation to assess model bias/ratios can often lead to ambiguous interpretations due to the Gaussian scaling required in such comparisons (Doornbos, 2012). Overall, there is no one standard test to determine the best model.

Therefore, in this study, the models' performance is evaluated on several metrics: difference in density ($\rho_{diff} = \rho_m - \rho_o$), density ratio ($\rho_{ratio} = \rho_o/\rho_m$), the standard deviation $\sigma$, Pearson correlation coefficient $R$, model bias $B$, and the model error standard deviation $E$. Subscripts $o$ and $m$ represent Swarm-C data and model estimated values, respectively. Two alike density distributions would have a $\rho_{ratio}$ of 1. Density ratio limits the range of values when looking at the difference between the estimated density and the observed density and also allows to highlight the magnitude of the proportional difference in a more straightforward manner.

Model bias $B$ and model error standard deviation $E$ are defined similarly to Elvidge et al. (2014):

$$B = \overline{m} - \overline{o}. \tag{2}$$

Equation (2) gives the $B$ as the difference between the mean of model estimates ($\overline{m}$) and mean of observations ($\overline{o}$). The $E$, which is also sometimes referred to as the centered pattern root-mean-square difference, is calculated as

$$E = \sqrt{\sigma_m^2 + \sigma_o^2 - 2\sigma_m\sigma_o R}. \tag{3}$$

The use of $E$ as a measure of model variation is useful especially in comparing distributions that are usually strongly correlated. However, strong correlation does not imply that the two distributions share the same amplitude of variation. The construction of $E$ that is linked to the standard deviations of the two distributions $m$ and $o$ allows the distinction of the differences in the amplitude variation of the distributions. If two distributions are identical, $E$ will be 0. $E$ is different from the root-mean-square difference that is commonly used in literature where the two will only be equal if the model (distribution/pattern) that is compared has zero bias with respect to the truth/observations as $E$ is the root-mean-square difference sans the mean (Elvidge et al., 2014; Taylor, 2001). This method is, however, limited to evaluate only the sampling variability in the model and not in the observations. The uncertainty in observations is not provided for the Swarm-C density product (used in this analysis), and therefore, the data are treated as the "truth" in the comparison.

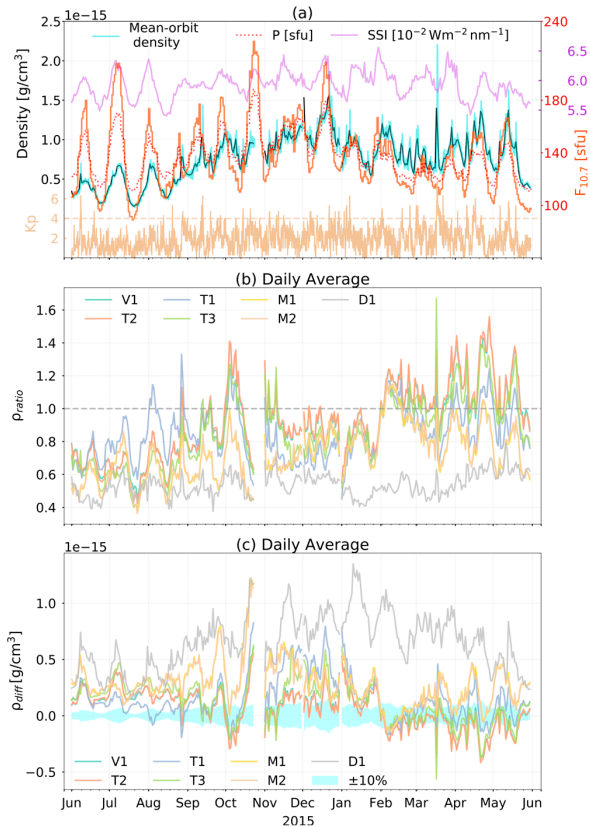

**Figure 1.** (a) Swarm-C density (daily average in black and orbital average in blue), 3-hr $Kp$, daily $F_{10.7}$, and $P$ solar flux (sfu = $10^{-22}$ W · m$^{-2}$ · Hz$^{-1}$), and daily SSI (solar spectral irradiance) in the 0.5- to 194.5-nm range. (b) Daily ratio of data/model ($\rho_{ratio}$) and (c) model-data difference ($\rho_{diff}$) for the runs considered in the study. Area within ±10% of data is shaded.

For a deeper investigation of the impact $Kp$ and $F_{10.7}$ (space weather parameters) on density estimates, a machine learning cross-validation model based on Pedregosa et al. (2011) was built. The cross-validation scheme attempts to predict the variability in a portion of the density distribution (validation set) by learning the response of density to space weather parameters in another portion of the respective distributions (training set). In this experiment, the entire time series was split into 100 parts and iterated over 100 times giving 99 training sets and one validation set for each prediction step. The folding value 100 to split the data set was empirically chosen as smaller (e.g., 10) or larger (e.g., 1,000) folding value turned out to be nonoptimal. The average of the resulting RMSE is considered in the results shown below. Note that these are the RMSE of the predicted values from the cross-validation scheme and not the RMSE for the different distributions compared as a whole. While each model calibrates space weather parameters differently, this test allows us to recognize the amount of relative influence each of the space weather parameters has on the observed variation in density.

## 4. Results

Figure 1 shows the solar and geomagnetic activities and the Swarm-C accelerometer-derived density during June 2014 to May 2015. The daily average (black) and orbital average (blue) of the Swarm-C accelerometer-derived density, $P$ index (red dashed), $F_{10.7}$ (red), the solar spectral irradiance (SSI; purple) in the 0.5- to 194.5-nm range and the 3-hr geomagnetic activity $Kp$ (brown) are shown in Figure 1a. Figures 1b and 1c display the daily averages of $\rho_{ratio}$ and $\rho_{diff}$ for V1 (green), T1 (blue), T2 (orange), T3 (yellow green), M1 (yellow), M2 (brown), and D1 (gray) model-data comparisons, respectively. The gaps in the line graphs in Figure 1 mark the data gaps, which are 23 October to 2 November, 30 November to 3 December, 21–22 December, and 31 December 2014. The shaded area in Figure 1c represents ±10% of the daily average of Swarm-C data.

Figure 1a shows the good correspondence of the Swarm-C data with the solar flux variations in both the proxies and measured SSI/EUV. The EUV data shown here are the postprocessed measurements from the solar extreme ultraviolet experiment instrument on board the Thermosphere Ionosphere Mesosphere Energetics and Dynamics satellite, which has an observation cadence of ∼3 min. While the variations of all three solar activity indicators (i.e., $P$, $F_{10.7}$, and SSI) are quite similar, there are some subtle differences, for example, the smooth EUV variations relative to the $P$ and $F_{10.7}$ during September–October and the nonproportionality in the amplitude differences. The higher-frequency variations in the orbit-averaged Swarm-C data show some similarities to the variations in the $Kp$, for example, in late August and mid-September 2014, and March 2015. Daily averages for solar activity and 3-hourly averages for geomagnetic activity are shown here to display those parameters on a scale comparable to the resolution available to the models. In other words, TIE-GCM, NRLMSISE-00, and DTM-2013 take in $F_{10.7}$ in the order of diurnally averaged and geomagnetic proxies in the order of hourly averaged. Interestingly, the variations in $\rho_{ratio}$ and $\rho_{diff}$ (Figures 1b and 1c) also seem to reflect the solar activity variations and the geomagnetic activity variations in Figure 1a. For example, the crest and trough locations in $\rho_{ratio}$ and the solar indices during July, as well as the sudden variations in $\rho_{ratio}$ and $\rho_{diff}$ in late August, coincide with significant changes in the *Kp*.

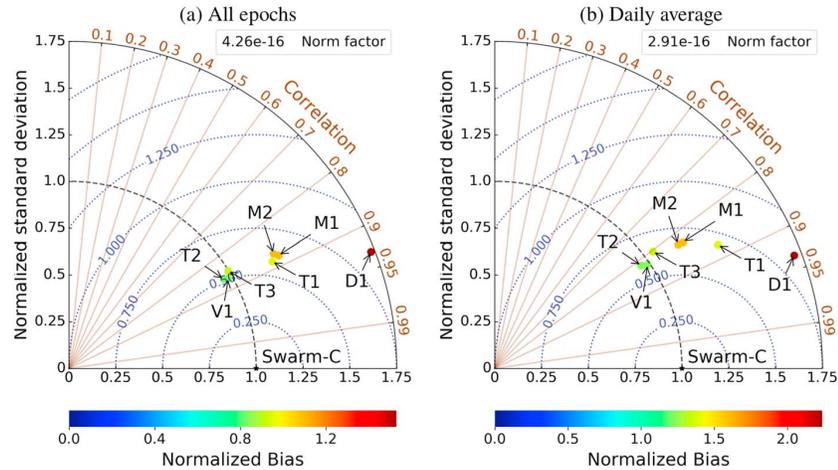

**Figure 2.** Modified Taylor diagram (Elvidge et al., 2014) showing the low-order statistics for the model scenarios introduced in Figure 1. The bias-free reference point is labeled as Swarm-C. The locus from the origin (0,0) bears the normalized standard deviation. The angle subtended by the standard deviation arc represents the correlation coefficient. The blue-dotted arc indicates the model error standard deviation. The model bias is shown on the color scale. The *Norm factor* is used to scale the relevant statistical parameters. (a) All epochs. (b) Daily average.

When comparing the performance of the individual model runs to the Swarm-C data in Figure 1, it can be observed that prior to September, the T1 run performed the best with $\rho_{ratio}$ close to 1 and the smallest overall values for $\rho_{diff}$. However, from September onward to about February 2015, the T2, T3, and V1 runs typically outperform the T1, M1, and M2 runs. D1's performance is under par throughout the time series. There is a notable performance improvement for all models from February onward where the physics-based model runs mostly fall within ±10% of Swarm-C density. During this period (February onward) T3 seems to outperform the rest.

The gradual transition from the midyear semiannual minimum to maximum in density is also visible in Figure 1a where the density in December–January is 2 to 3 times higher than in June.

Figures 2a and 2b summarize the statistical results of the comparisons between the model runs and the Swarm-C data for all epochs and the daily averaged Swarm-C data, respectively. The low-order statistics displayed in Figure 2 in a modified Taylor diagram (MTD) are the $\sigma$, $R$, $E$, and $B$ (Elvidge et al., 2014; Taylor, 2001). The black star labeled "Swarm-C" indicates the bias-free hypothetical truth point to which the rest are referenced. At this point, $\sigma$ equals the normalizing factor, $R$ equals 1, and $E$ equals 0. The radial distance from the origin indicates the $\sigma$, normalized to the standard deviation of the Swarm-C data (black dashed line). $R$ is delineated on the outer edge of the standard deviation arc. The dot marking the model's position on the diagram also represents the normalized $B$ given in the color bar underneath. $B$ is calculated as per equation (2). The blue-dotted arc indicates $E$ as per equation (3). The absolute values for $\sigma$, $E$, and $B$ are obtained by multiplying by the respective normalization factors indicated in Figure 2.

Figure 2 shows that comparing model performance for averaged values versus per each epoch leads one to draw different conclusions about the models' performance. However, it is clear that TIE-GCM outperforms NRLMSISE-00 in all four metrics and DTM-2013 in all but $R$ in both data comparisons. The complete statistics corresponding to all seven runs are given in Appendix A1 along with an MTD for mean orbital performance. A noteworthy result of Figure 2 is that applying a temporal average on the data (i.e., taking the daily average of the 10-s Swarm-C data) does lower the overall $\sigma$, $E$, and $B$ for all models and significantly changes $R$ for all models except T1 and D1.

It can be seen that in general DTM-2013, NRLMSISE-00, and TIE-GCM are positively biased during the analysis period. Elvidge et al. (2016) also observed that NRLMSISE-00 and TIE-GCM have a tendency to be positively biased in a study comparing the two models to CHAMP accelerometer-derived densities.

Overall, the results in Figure 2 show that the V1, T2, and T3 runs that employ a constant $K_{zz}$ coefficient bear the closest resemblance to the Swarm-C data—the lowest $E$ and $B$. Interestingly, the T1 run, which employed a seasonally varying $K_{zz}$ coefficient, performed significantly worse in terms of standard deviation. The statistical results from the M1 and M2 runs are all very similar to each other, indicating that the different NRLMSISE-00 runs had a little overall impact on their comparisons against the Swarm-C data. DTM-2013 model run although has recorded the best correlation coefficient has the worst standard deviation and model bias compared to the rest of the model runs.

A clear correlation between $\rho_{ratio}$ and $\rho_{diff}$ to $Kp$ and EUV/$F_{10.7}$ is apparent in Figure 1. The models use, for example, the imperfect proxies $Kp$ and $F_{10.7}$ for geomagnetic and solar activities, respectively. To further investigate the relationship between these proxies and model error, Figure 3 presents a composite of mean orbit bivariate distributions of $\rho_{diff}$ as a function of $Kp$ (left column) and $F_{10.7}$ (right column). The vertical axes in Figures 3a–3n show the model-data density difference ($\rho_{diff}$) for each model run. The positive and negative values on the vertical axes indicate the model overestimate and underestimate, respectively. The green dots show the scatter distribution, while the contour lines are drawn in the order of increasing color intensity to display the bivariate probability density distribution. The separation of contour lines is directly related to the mean integrated squared error, which is used for the estimation of the bandwidth for the kernel density. The linear regression fit is shown by the blue line, and its slope is displayed as $S$. The Pearson correlation coefficient is displayed as $R$. The axes limits for each panel are kept the same for ease of comparison.

The left column of Figure 3 shows that for the most part the geomagnetic activity was low ($Kp < 5$) during the analysis period. The $\rho_{diff}$ distributions in Figures 3c and 3k (T1 and M2) show no discernible increasing or decreasing linear pattern with respect to $Kp$. While V1, T2, and T3 show a weak negative correlation empirical model, runs M1 and D1 show a weak positive correlation. The scatter distributions suggest no significant linear relationship between model-data difference and increasing geomagnetic activity, bar D1.

On the other hand, the right column of Figure 3 shows a clear moderate increase in $\rho_{diff}$ for all model runs with increasing solar activity. T1 has the highest correlation, and T2 and D1 have the lowest correlation. Interestingly, the only difference between T1 and T2 is that the former is driven with daily varying $K_{zz}$ and the latter with constant $K_{zz}$. The slope values indicate that the rate of change in $\rho_{diff}$ with increasing solar activity is small relative to the average density. This linear relationship indicating that model performance degrades with increasing $F_{10.7}$ may be somewhat exaggerated because the magnitude of the density in the thermosphere is increased in general during high solar activity.

Figure 4 shows the deviation of density ratio from the ideal ($\rho_{ratio} = 1$) for a given level of solar and geomagnetic activity. The corresponding model run is labeled at the top of each panel. The color scale for panels (a)–(g) is centered at 0, where the positive and negative values indicate the model overestimate and underestimate, respectively. The representation in Figure 4 is derived by forming $Kp$ from 1 to 8 and $F_{10.7}$ from 80 to 220 sfu into an $8 \times 8$ matrix of equal bin size where the average of each bin is taken as the representative value. The underlying data set corresponds to the orbit-averaged densities. The number of events per bin is shown in Figure 4h.

It is clear in Figure 4 that the physics-based model outperforms the empirical models at low to moderate geomagnetic activity levels across the $F_{10.7}$ axis. TIE-GCM's tendency to overestimate density at higher $Kp$ and moderate $F_{10.7}$ (~110–135 sfu) that was apparent in Figure 3 is further accentuated in this figure. That tendency, however, is somewhat subdued by the inclusion of the day-of-year-dependent $K_{zz}$. It is understandable from Figure 4 that NRLMSISE-00 experiences a performance degradation at low and high $F_{10.7}$ but M1 performs slightly better than the rest of the model runs when $Kp$ is highest, which also happens to be the bin where D1 records its best performance. The performance of D1 is seemingly indifferent to the strength of space weather proxies. As indicated in Figure 4h, the number of data points belonging to each bin is not equal throughout and much less in the high end of the spectrum than in the middle. Therefore, the performance statistics from the high end of $Kp$ and $F_{10.7}$ spectrum are not weighted the same as the low to moderate activity levels.

Figure 5 offers a neutral comparison of model and data behavior (i.e., not in terms of performance against data) with respect to select space weather parameters. The models' inclination toward space weather proxies is investigated in Figure 5 using a machine learning algorithm that peruses the density distributions at various intervals. The process is carried out by forcing predictions entirely based on EUV flux, $F_{10.7}$, and $Kp$ (referred to as "feature" distributions in the machine learning parlance), taking each distribution into account separately and a combination thereof as shown in Figure 5 using color-coded labels. Figure 5 shows the RMSE of the predicted density against the corresponding density distribution, which includes the Swarm-C accelerometer-derived density displayed as SW.

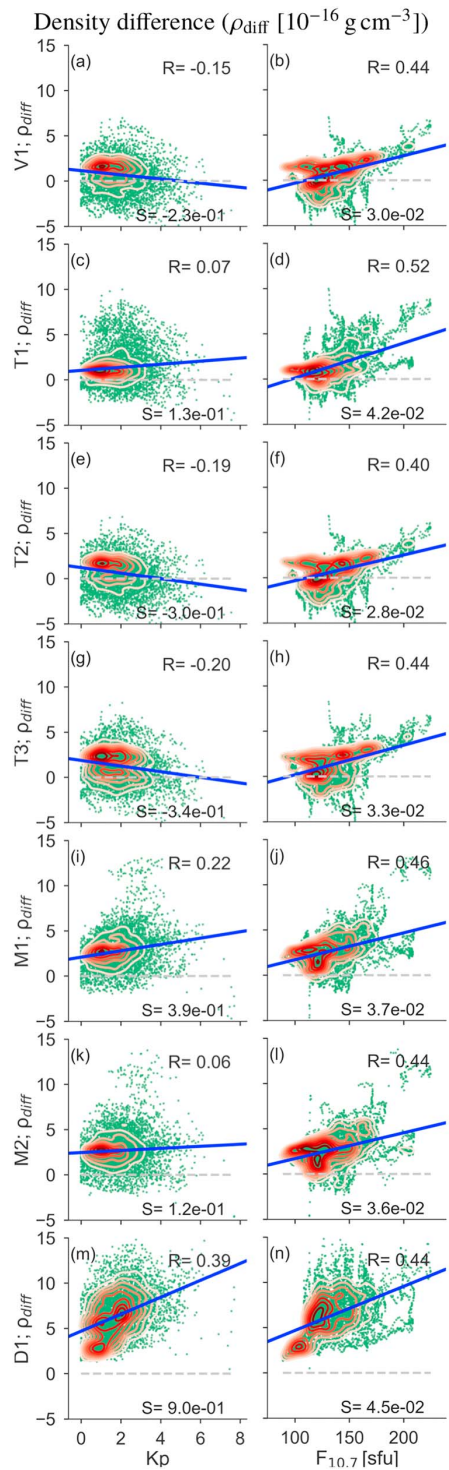

**Figure 3.** (a-n) The correlation of the $Kp$ and $F_{10.7}$ with $\rho_{diff}$ of the specified model runs considering the Swarm-C's mean orbital variability. The green dots show the scatter distribution, while the contour lines represent the bivariate probability density distribution. The slope of the blue linear regression fit line is displayed as $S$. $R$ is the Pearson correlation coefficient. The broken line provides the comparable Swarm-C reference.

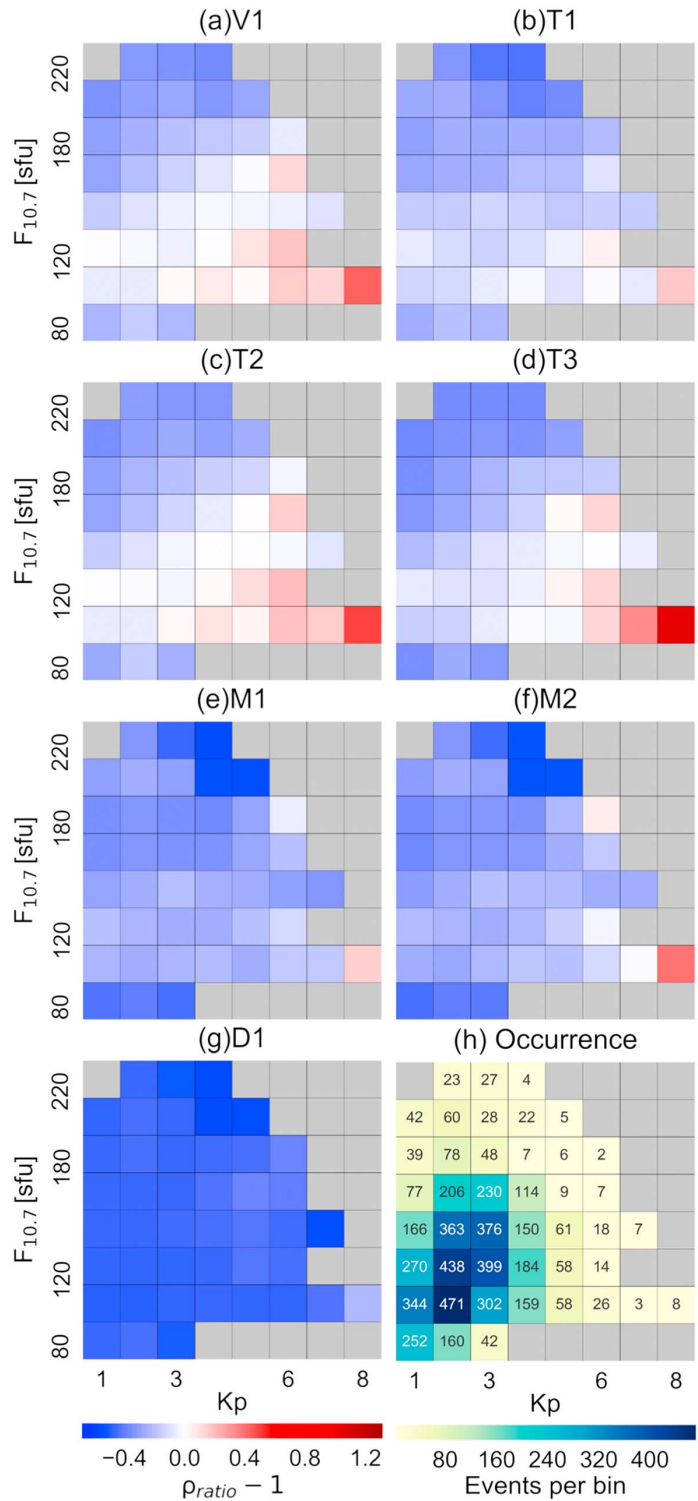

**Figure 4.** Performance of the specified model runs at different solar and geomagnetic activity levels displayed as deviation from the ideal density ratio ($\rho_{ratio}$) corresponding to Swarm-C mean orbital density. (a–g) The average $\rho_{ratio} - 1$ per bin is represented by the blue-red color scale. (h) The number of events per bin.

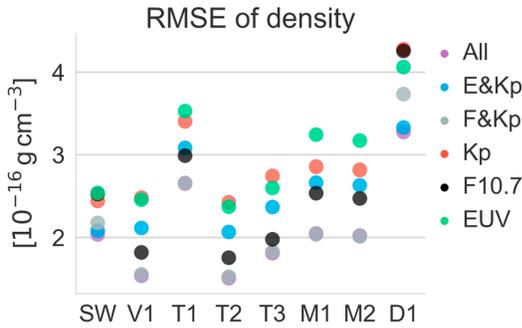

**Figure 5.** The 1-D graph for the root-mean-square error (RMSE) results from the machine learning cross-validation scheme. SW stands for the Swarm-C accelerometer-derived density, and other column labels are for the different model runs. Feature labels represent the distribution used in the training. EUV, F10.7, and Kp correspond to predications made by using EUV flux, $F_{10.7}$, and $Kp$ distributions, respectively, in the training independent of each other. Combined distributions used for training are as follows: F&Kp = $F_{10.7}$ and $Kp$; E&Kp = EUV and $Kp$; All = EUV, $F_{10.7}$, and $Kp$. EUV = extreme ultraviolet.

The variations in all the representations of thermospheric density in Figure 5 show low RMSE when trained upon the combined variations in EUV flux, $F_{10.7}$, and $Kp$ (purple). The Swarm-C data (SW) show a slightly higher inclination to the combined variations in E&Kp (blue) than to F&Kp (gray). Interestingly, the machine learning exercise demonstrates how the TIE-GCM and NRLMSISE-00 models are inherently far from E&Kp variations compared to F&Kp or F10.7 (black). This is not surprising as both the models use $F_{10.7}$ as the solar proxy and therefore inherently unaware of the variations in EUV flux. DTM-2013 that uses $F_{30}$ as the solar proxy, on the other hand, demonstrate a better inclination to E&Kp than to F&Kp. DTM-2013 also shows higher RMSEs when trained on each feature distribution separately. All density distributions record poor RMSEs when trained solely on EUV flux.

It is also clear in Figure 5 that when just the $Kp$ distribution is used in the training, the RMSE is relatively high. As expected, when all distributions are featured in the training set, the RMSE is significantly improved. The relative differences between the RMSE values in each column demonstrate how dominant each feature variable in describing the variations in the respective density distributions.

The columns representing the TIE-GCM runs in Figure 5 have the lowest RMSEs compared to the other distributions. The significant difference between T1 and the other three TIE-GCM runs that was apparent in the previous figures is also present in the predicted RMSE values. The most explicit feature among the TIE-GCM runs is the relative proximity of predictions based on $F_{10.7}$ to that of F&Kp in comparison to such proximity among the other distributions.

Figures 1–5 indicated that the physics-based model is somewhat superior compared to the two empirical models during the analysis period. In Figure 6, the physics-based model's performance at the high latitudes is further investigated. The two MTDs in Figure 6 provide some insights into TIE-GCM behavior during Swarm-C orbit segments within 50° from the poles. It can be seen that performance in the Southern Hemisphere

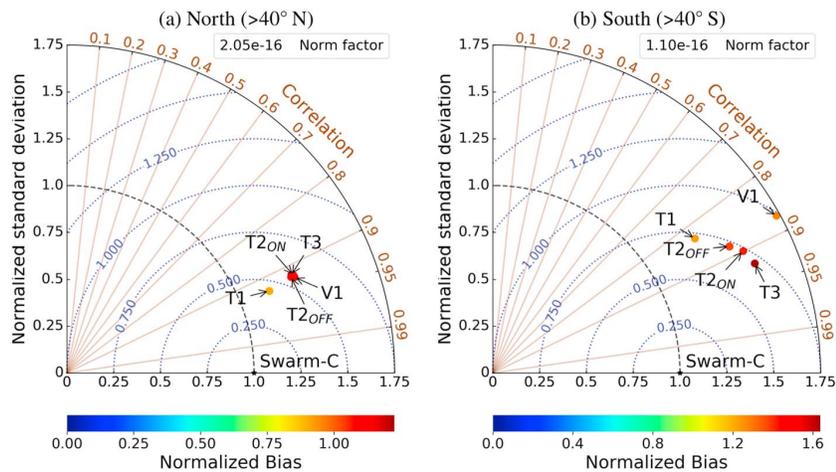

**Figure 6.** Modified Taylor diagram (as described in Figure 2) showing the low-order statistics for select Thermosphere-Ionosphere-Electrodynamics General Circulation Model model runs comparing the influence of He on model performance at high latitudes during June–July 2014. Subscripts ON (OFF) for T2 indicate He included (not) in model chemistry as a major constituent. He is not among the major species accounted for V1. The significant difference among the V1 and T2 runs in (b) maybe due to the wintertime He-bulge formation. (a) North (>40°N). (b) South (>40°S).

(Figure 6b) is poorer than the Northern Hemisphere (Figure 6a). In other words, the relative standard deviation and model bias is higher for the Southern Hemisphere distributions than their counterparts in the Northern Hemisphere. The correlation coefficients have also degraded in general.

Two runs with T2 are shown in Figure 6: $T2_{ON}$ and $T2_{OFF}$. $T2_{ON}$ ($T2_{OFF}$) represent the simulation with (without) helium being added to the model chemistry. Figure 6b clearly shows that TIE-GCM version 2 performs better than the previous version in terms of standard deviation and model error standard deviation. Comparing the differences between V1, $T2_{ON}$ and $T2_{OFF}$ indicates the impact of helium on the model performance and the improvement to neutral density product in the latest version. $T2_{ON}$ in Figure 6b shows a slightly higher correlation than $T2_{OFF}$, which is the expected behavior in the winter hemisphere with a higher concentration of helium than the summer hemisphere (Mayr et al., 1978; Reber & Hays, 1973; Sutton, 2016).

## 5. Discussion

The model performance with respect to both $\rho_{diff}$ and $\rho_{ratio}$ is visualized in Figure 1. One of the main highlights from our results is the importance of not relying on one metric in model-data comparisons. The two metrics, $\rho_{diff}$ and $\rho_{ratio}$, essentially decide the same fact (i.e., if the model overestimated or underestimated). While all three models typically overestimate density compared to Swarm-C data, TIE-GCM underestimated density mostly during periods of increased geomagnetic activity (i.e., $Kp > 4$; see Figure 1). However, for example, in June and December 2014 when $Kp$ upsurged, a corresponding lower-than-data estimate of density among the TIE-GCM runs is not observed. In fact, V1 and T2 runs performed mostly within $\pm 10\%$ of Swarm-C density in December 2014.

NRLMSISE-00 also underestimated density briefly in February and April 2015. In agreement with Doornbos's (2012) discussion about $\rho_{diff}$ and $\rho_{ratio}$, it can be seen that model response at these instances is clearly magnified in Figure 1b more than in Figure 1c. In this comparison against other dependent distributions (e.g., $F_{10.7}$ or $Kp$), Figure 1 shows that $\rho_{ratio}$ highlights model performance features better than $\rho_{diff}$. Also, for example, the $\rho_{ratio}$ for the density spike in late August that correspond to a surge in $Kp$ from 1 to 4 is more noticeable than its counterpart $\rho_{diff}$ in Figure 1. Furthermore, the amplitude differences inherent within different distributions may also sometimes lead to ambiguous interpretations in an evaluation based on $\rho_{diff}$ and $\rho_{ratio}$. For example, Figure 1 shows that NRLMSISE-00-$\rho_{diff}$ in late October is more than twice than in June. Still, $\rho_{ratio}$ for the same reveals that model performance did not change as much indicating the usefulness of $\rho_{ratio}$ in model evaluation.

Figures 1b and 1c reveal that DTM-2013 never underestimated density during the entire analysis period. More importantly, Figure 1b shows that the performance of DTM-2013 did not vary as much as the other model runs. It can be seen that there is a slight increase in DTM-2013's performance in the latter end of the time series. Unlike the other model runs, DTM-2013 never fell within the range of $\pm 10\%$ of Swarm-C density.

The MTDs shown in Figure 2 present the gist of the model-data comparison concluding that on average TIE-GCM outperforms NRLMSISE-00 and DTM-2013 for the given model runs. As mentioned earlier, it is common to consider daily or hourly averaged values in similar comparisons. The MTDs reveal that how the underlying distribution is treated affects model performance rankings and conclusions derived from such a comparison. Our MTDs show the average results for the entire analysis period. This comparison could be extended to look at performance based on various other attributes such as season, solar and geomagnetic activity level, altitude, and geographic location, which is beyond the ambit of this paper. An additional MTD considering orbit-averaged densities is given in Figure A1 for comparison.

As presented in Figures 3 and 4, the correlation of geomagnetic activity to model performance is an important metric to consider as it is one of the main drivers of density. In other words, the manner in which the geomagnetic proxies are applied in the model affects the model performance. The $Kp$'s role in TIE-GCM contributes to the calculation of high-latitude energy input to the thermosphere system via Joule heating and particle precipitation. When comparing the effect of ion convection pattern used in TIE-GCM(H), clearly the overall performance of TIE-GCM(H) seems to be similar to TIE-GCM(W) during the studied period. The time series in Figure 1, however, shows that TIE-GCM(W) performs poorer than TIE-GCM(H) during the second half of 2014 and better from February onward. Figure 6 showed that at the high latitudes where the influence of the ion convection model is strongest, the difference between Heelis et al. (1982) and Weimer (2005) models is more pronounced only in the Southern Hemisphere.

Wu et al. (2015) showed that in resolving thermosphere winds, the Weimer (2005) model performs better during storm times than the Heelis et al. (1982) model. Interestingly, Figure 1 shows a few instances where TIE-GCM(W) estimated lower densities than TIE-GCM(H) coinciding with instances of enhanced $Kp$. For example, T3 (i.e., driven by Weimer, 2005, model) and T2 (i.e., driven by Heelis et al., 1982, model) in Figure 1 during the few instances where $Kp$ exceeded 6. It also appears that T3's performance against T2 is poorest during mid-November to December (see Figure 1). The time series is not adequate enough to suggest a seasonal dependency attached to the model performance at Swarm-C altitude when driven with Weimer (2005) model or Heelis et al. (1982) model.

The role of geomagnetic activity in NRLMSISE-00 can be summed up as follows. Regardless of the input $Ap$, NRLMSISE-00 sets $Ap = 4$ by default for altitudes below 80 km (Picone et al., 2002). The NRLMSISE-00 model determines an upper limit for temperature based on $F_{10.7}$ and $Ap$. Then based on the underlying data set, linear, exponential, and cross terms with $Ap$ are drawn for each of the represented constituents to fit in the temperature formulation as a function of diffusive equilibrium (Picone et al., 2002). It is a complicated process involving the determination of a large number of nonzero coefficients and correction parameters through model calibration, which is expected to absorb some of the deficiencies in the correlation between input proxies. Studying the differences between M1 and M2 reveals that NRLMSISE-00's performance seems to be not particularly affected much by the extent of the history of $Ap$ fed into the model (see Figures 1, 2, and Appendix A).

The differences between the two methods of driving NRLMSISE-00 (e.g., M1 and M2) are anticipated to change with geomagnetic activity and the sunspot cycle due to the effect of geomagnetic activity on density. Therefore, the different methods of driving NRLMSISE-00 may especially impact the results at high latitudes during high geomagnetic activity. To further elucidate the effects of driving NRLMSISE-00 with the daily average of 3-hr $Ap$ or the full required history of $Ap$, results from two previous studies are consulted, Elvidge et al. (2016) and McLaughlin et al. (2013), and reported in Appendix B.

The first example, given in Figure B1, corresponds to the CHAMP accelerometer-derived density (see Sutton, 2009) during the period 28 August to 1 September 2009 (i.e., solar minimum) that included a geomagnetic storm with $Ap$ striking a high of 67. The comparison presented in Figure B1 is similar to "Test 1" in Figures 2 and 3 in Elvidge et al. (2016). Figure B1a shows model estimates and CHAMP data at a temporal resolution of approximately 45 s. The $\sigma$, $R$, $E$, and $B$ for the corresponding hourly averaged distributions are given in Figure B1b. The significant enhancement gained in NRLMSISE-00's performance when geomagnetic activity is introduced as a matrix of seven periods of anterior magnetic activity is evident.

The second example, given in Figure B2, compares CHAMP and GRACE-A accelerometer-derived densities during 26–27 September 2007 with the above mentioned two different geomagnetic activity input methods. The comparison presented in Figure B2 is similar to the CHAMP and GRACE-A comparison presented in Figure 5 in McLaughlin et al. (2013). The model-data comparison is performed at the original data resolution. It is abundantly clear that NRLMSISE-00's performance, when driven with the full required history of $Ap$, is superior to that of the daily average $Ap$ during the short periods belonging to solar minimum shown in Figures B1 and B2.

Menvielle and Berthelier (1991) describe the computation process of $Ap$, $Kp$, and $am$ indices and note that $am$, which is used in the DTM-2013, is a better indicator of small perturbations during low geomagnetic activity periods as well as high perturbations in geomagnetic activity than $Kp$. $Ap$ and $Kp$ essentially encompass the same information, and different levels are weighted the same. $Ap$ is the equivalent linear scale of the quasi-logarithmic $Kp$ scale.

As for the correlation of solar activity to model performance, the similarity in the trends between the solar proxies shown in Figure 1a and, for example, the trends in Figure 1b shows that insolation determines and dominates the model estimated densities at Swarm-C's altitude. Figure 1a highlights instances where $P$ and $F_{10.7}$ demonstrate a trend that is different to EUV, for example, the steep downward trend in $F_{10.7}$ in late November that is reflected in neither EUV nor the Swarm-C data. Another example is the upward trend in EUV in April that is not seen in $F_{10.7}$. It can be easily deduced that the differences in variation between EUV versus $P$ and $F_{10.7}$ only become more significant on even smaller timescales. This can result in erroneously resolved energy input to the thermosphere system by the EUVAC solar proxy model inside TIE-GCM (Solomon & Qian, 2005). Further, the machine learning cross-validation scheme reveals the isolated systematic bias of each model run to space weather parameterization, where TIE-GCM and NRLMSISE-00 are more inclined toward variation in $F_{10.7}$ flux than the DTM-2013.

Out of the four different TIE-GCM runs, it is evident that the introduction of day-of-year-dependent $K_{zz}$ makes a significant impact on the model estimates. In other words, the inclusion of empirically derived eddy diffusivity pushes the model estimates closer to Swarm-C data in months following June solstice and farther away post-September equinox until February (see T1 in Figure 1). Compared to T1, a significant performance degradation for the runs with constant $K_{zz}$ happens around July and August (e.g., T2 in Figure 1). The time series is not adequate enough to suggest that the model runs with constant $K_{zz}$ have a seasonal dependency on performance at Swarm-C altitude. As shown in Qian et al. (2009, Figure 10), the constant $K_{zz}$ performs poorly around July (i.e., semiannual minimum), and our results show on average, driving TIE-GCM with constant $K_{zz}$ seems to be more effective in this comparison. Nonetheless, from a modeling perspective, these trends suggest that it is beneficial to include daily varying $K_{zz}$ in density forecasts during June–September.

A notable distinction between the T1 run and the other three TIE-GCM runs is apparent throughout. Some insights into how the physics-based model responds to variations in geomagnetic activity when the lower and middle atmospheric forcing in the form of $K_{zz}$ is varied daily is found in the results presented here. Figure 3c shows no discernible correlation between T1's performance and $Kp$. Figure 4 shows that T1's performance at high $Kp$ and low $F_{10.7}$ is better than V1, T2, and T3. The results from the machine learning cross-validation scheme (see Figure 5) revealed that T1's relative inclination to variations in $Kp$ is slightly better than it is for EUV flux, whereas the inverse is true for the other three TIE-GCM runs. However, the relative inclination to other distributions presented in Figure 5 (i.e., except Kp and EUV) is similar for all the TIE-GCM runs. The distinction between T1 and the other TIE-GCM runs connotes that variable eddy diffusivity imparts on $Kp$ forcing in the physics-based model. T1, which included helium in the model chemistry by default, also performed the best at orbital segments corresponding to the Northern Hemisphere but performed slightly poor in the Southern Hemisphere (see Figure 6). Such hemispherical differences highlight the vulnerability of TIE-GCM driven with the empirically formulated daily varying $K_{zz}$ during those segments along the Swarm-C orbit.

$K_{zz}$'s primary function is to emulate the semiannual oscillation observed in the thermosphere and ionosphere. The $K_{zz}$ coefficient used in TIE-GCM contains more than just the short wavelength eddy diffusion (i.e., turbulent mixing of breaking gravity waves at a smaller scale than model's grid resolution) and is representative of other lower atmospheric disturbances originating below TIE-GCM's lower boundary such as tidal dissipation (Jones et al., 2017). Using the thermosphere-ionosphere-mesosphere electrodynamics general circulation model, Jones et al. (2017) show how lower and middle atmospheric forcing influences the semiannual oscillation in the thermosphere and manages to produce these effects without resorting to $K_{zz}$. As self-consistent gravity wave parameterization is not integrated into TIE-GCM, further analysis with modeling these force terms in the Navier-Stokes equations will be required to forgo the empirical representation of $K_{zz}$.

T2 and V1 runs where the only difference is the model version, produced very similar results. Although lack of helium is not expected to introduce severe errors in total density at Swarm-C's altitude under solar maximum conditions, Figure 6 showed that T2 is at least slightly better than V1 in the winter hemisphere during June–July 2014. This indicates that the improvements made to the model chemistry have an impact even at this altitude. Sutton et al. (2015) show that at an altitude of 415 km, helium contributes about 10–15% in June solstice and 20–25% in December solstice to total density under 2008 solar minimum conditions. Sutton et al. (2015) also claim the contribution from helium under solar minimum conditions to be between 100% and 200% during solstice at an altitude of 500 km. The number densities of lighter species such as helium increase with altitude and thus also increase the associated molecular diffusion. The physical mechanism of the tendency of, for example, helium to concentrate in the high latitudes of the winter hemisphere is attributed to vertical advection and horizontal transport assisted by the thermosphere wind system (e.g., Mayr et al., 1978; Reber & Hays, 1973; Sutton, 2016). Our results show the implication of accounting for helium compared to highly precise accelerometer-derived densities. However, no noticeable difference was observed for northern winter hemisphere during December–January (not shown).

Results from the machine learning cross-validation scheme in Figure 5 show that TIE-GCM and NRLMSISE-00 are less inclined to combined variations in EUV flux and $Kp$ than in $F_{10.7}$. DTM-2013, on the other hand, shows an increased inclination to the combined variations in EUV flux and $Kp$. Figure 5 offers some insights into inherent characteristics of individual density distributions (model and data) to major space weather parameters. This behavior of model density distributions on the machine learning exercise maybe linked to the proxy information available to each model. The Swarm-C data display an RMSE response similar to the NRLMSISE-00. The RMSEs in data are also another way to illustrate that the variation in thermospheric density is not a simple function of EUV, $F_{10.7}$, and $Kp$ alone.

As mentioned previously, uncertainty in the Swarm-C data is not considered in this study. As Siemes et al. (2016) point out, uncertainty in the accelerometer measurements and biases induced due to derivation technique are present in the data. One of the important steps in extracting density in the technique used for Swarm-C is to estimate background winds using the Drob et al. (2008) empirical horizontal wind model. Drob et al. (2015) highlight the substantial disagreements in Drob et al. (2008) compared to observations. The Drob et al. (2008) model ignores the vertical winds and only represents geomagnetically quiet conditions ($Kp < 3$). The Drob et al. (2008)-DWM07 component accounts for the winds during geomagnetically disturbed periods. Furthermore, even the improved version of the Drob et al. (2008) model lacks solar flux parameterization and among others, there are also issues with describing quiet time high-latitude circulation patterns (Drob et al., 2015).

The errors due to background winds are at least an order of magnitude less in comparison to spacecraft along-track velocity vector (Doornbos, 2012). Therefore, its effect on the derived density product may not be of consequence compared to other more significant errors due to accelerometer calibration. Using physical winds in the accelerometer-density derivation process has not been attempted, so far, to our best knowledge. Winds resolved by self-consistent physical models may help mitigate some of the fundamental issues inherent in empirical winds. Especially in a nowcast/forecast setting, numerically computed winds will eliminate the dependency on statistically averaged winds during disturbed space weather conditions, thereby improving the reliability of the derived density product.

## 6. Summary and Conclusions

The first evaluation of the Swarm-C accelerometer-derived densities using the two latest versions of TIE-GCM and two widely used empirical models, NRLMSISE-00 and DTM-2013, was presented. This extended satellite epoch-wise comparison for a continuous 6 months, as opposed to studying specific events, proved useful in evaluating the models' fidelity for forecasting applications. TIE-GCM outperforms the empirical models in almost all the metrics used in the comparison. The use of data to model ratio and difference as comparison metrics is discussed. While models show good agreement with the data in both metrics, the ratio is a better performance indicator for comparisons with solar and geomagnetic activities. A similar conclusion is made in Doornbos (2012).

The results show the complex interconnectedness of solar activity and geomagnetic activity on model performance in terms of estimating density. This case generally agrees with earlier work concerning model performance during different levels of solar and geomagnetic activity (e.g., Emmert et al., 2014; Solomon et al., 2011).

The impact of two key boundary conditions, eddy diffusion and ionospheric convection, on TIE-GCM's density estimation was analyzed. TIE-GCM shows a strong bias to the specification of the lower atmospheric eddy forcing. Over the period of nearly 12 months, model runs corresponding to TIE-GCM(W) performed similarly to TIE-GCM(H). The seasonal and geographical differences in performance between TIE-GCM(W) and TIE-GCM(H) were discussed.

The sensitivity of NRLMSISE-00 model to the specification of geomagnetic forcing was also analyzed. During the analysis period where the overall geomagnetic activity was quiet, driving NRLMSISE-00 with just the daily average of $Ap$ did not seem to have a contrasting effect compared to driving with seven histories of anterior magnetic activity. Using additional examples (given in Appendix B), replicating previous studies showed that specifying only the daily average $Ap$ significantly degrades the model performance. As shown in Solomon et al. (2011), depriving the NRLMSISE-00 of complete geomagnetic history results in larger residuals. Picone et al. (2002) note the model's frailty at high latitudes and high geomagnetic activity due to lack of observational data. The results from this study allude to the fact that the NRLMSISE-00 model does not capture the variations in shorter timescales associated with Swarm-C at high solar and geomagnetic activity levels. The results from this study also showed that DTM-2013 is not as successful as either NRLMSISE-00 or TIE-GCM in estimating density along the Swarm-C orbit.

The cross-validation scheme proved useful in demonstrating characteristics inherent to each model run. Our overall results also demonstrate that Swarm-C data are suitable for model validation and scientific study. The model performance evaluation could further be improved if the derivation-noise pertaining to Swarm-C data can be provided to the user.

Although TIE-GCM is computationally intensive compared to its empirical counterparts, the results shown here demonstrate that it has the potential to be utilized in density forecasting applications. In addition to forecasting density itself, physical models can also provide the winds required to derive density from accelerometer data. Nevertheless, it is fundamentally important to identify the discrepancies between the models and data. In this regard, state-of-the-art data assimilation techniques equip the self-consistent physical models with "self-healing" capabilities whereby the systematic and inherent model biases are corrected as the forecasting progresses to reflect the changes in the real-world observations. This aspect is particularly important to satellite orbit prediction and collision avoidance in LEO.

## Appendix A: Results From the Model Performance Evaluation

In Table A1, the low-order statistics considering the complete Swarm-C density distribution (i.e., all epochs) and the respective daily average denoted by the superscript a are presented. An MTD similar to Figure 2 but considering the orbit-averaged Swarm-C density distribution is presented in Figure A1.

**Table A1**
*Low-Order Statistics for the Ensemble of Model Runs*

| Run | Std | Std[a] | $R$ | $R$[a] | Model bias | Model bias[a] | ErrStd | ErrStd[a] |
|-----|-----|--------|-----|--------|------------|---------------|--------|-----------|
| V1 | 4.175E−16 | 2.876E−16 | 0.870 | 0.826 | 0.678E−16 | 0.690E−16 | 2.156E−16 | 1.706E−16 |
| T1 | 5.230E−16 | 3.967E−16 | 0.885 | 0.874 | 1.583E−16 | 1.615E−16 | 2.465E−16 | 2.004E−16 |
| T2 | 4.071E−16 | 2.784E−16 | 0.866 | 0.818 | 0.605E−16 | 0.616E−16 | 2.168E−16 | 1.723E−16 |
| T3 | 4.251E−16 | 3.060E−16 | 0.852 | 0.804 | 1.280E−16 | 1.306E−16 | 2.315E−16 | 1.875E−16 |
| M1 | 5.423E−16 | 3.523E−16 | 0.880 | 0.830 | 2.822E−16 | 2.827E−16 | 2.627E−16 | 1.964E−16 |
| M2 | 5.340E−16 | 3.442E−16 | 0.872 | 0.828 | 2.756E−16 | 2.760E−16 | 2.644E−16 | 1.929E−16 |
| D1 | 7.370E−16 | 4.977E−16 | 0.933 | 0.935 | 6.521E−16 | 6.517E−16 | 3.726E−16 | 2.478E−16 |

*Note.* Each run is as defined in section 3. Std is the standard deviation. ErrStd is the model error standard deviation. Except for Pearson correlation coefficient $R$, all other metrics have the dimensions g/cm³.
[a]The daily average of the corresponding all epochs distribution.

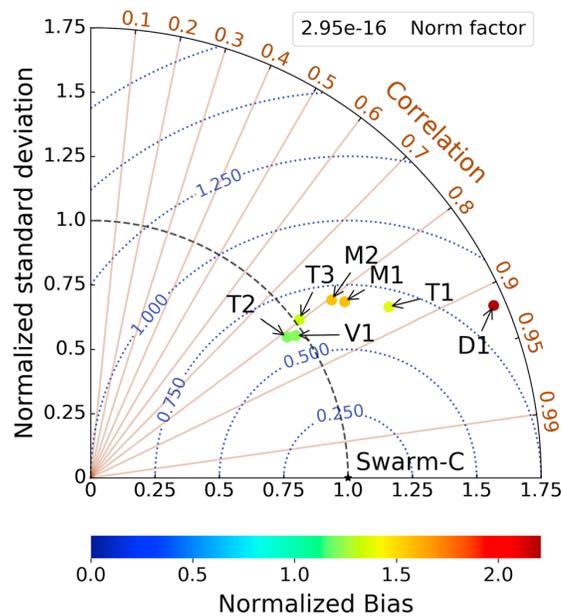

**Figure A1.** A modified Taylor diagram (Elvidge et al., 2014) similar to Figure 2 but for the orbit-averaged density distributions. T3 has the lowest standard deviation and lowest correlation with respect to Swarm-C density distribution.

## Appendix B: Effects of the Histories of Geomagnetic Activity on NRLMSISE-00

The examples given in Figures B1 and B2 demonstrate that NRLMSISE-00's performance can be significantly improved by driving the model with the full required history of $Ap$.

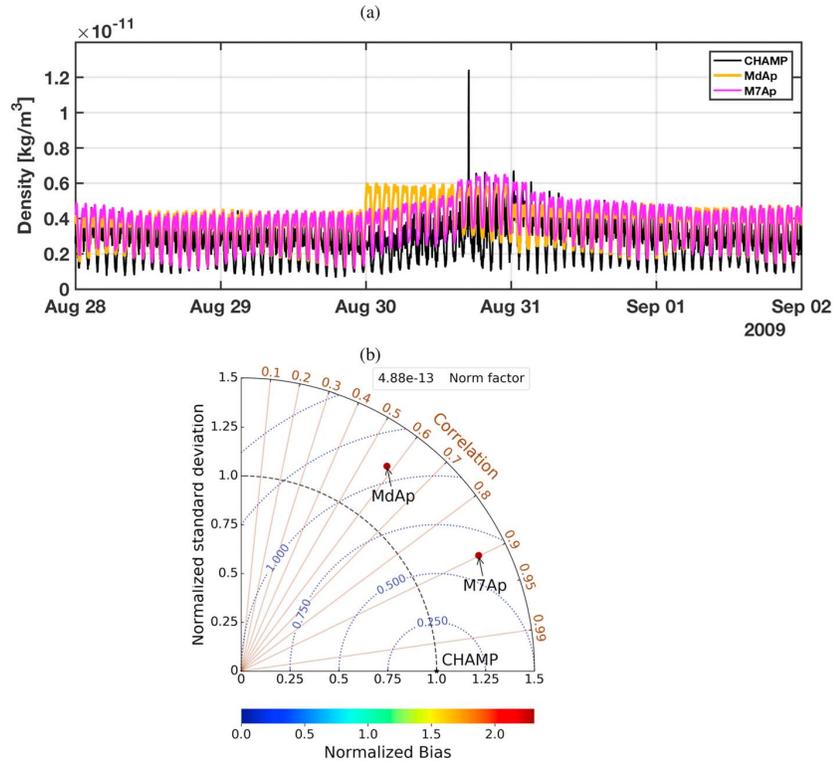

**Figure B1.** (a) Epoch-wise comparison of NRLMSISE-00 model estimates to CHAMP accelerometer-derived densities. MdAp run is driven with daily average of 3-hr $Ap$, and M7Ap is driven with the full required history of $Ap$. The spike on CHAMP data on 30 August corresponds to a geomagnetic storm that occurred between 15:00 and 18:00 UT. Average orbital height of CHAMP satellite during this period was 325 km. (b) A modified Taylor diagram (Elvidge et al., 2014) similar to Figure 2 but for the model runs in (a) considering the hourly average of the distributions. The significant enhancement gained in NRLMSISE-00's performance when geomagnetic activity is specified as per M7Ap is evident. CHAMP = Challenging Minisatellite Payload.

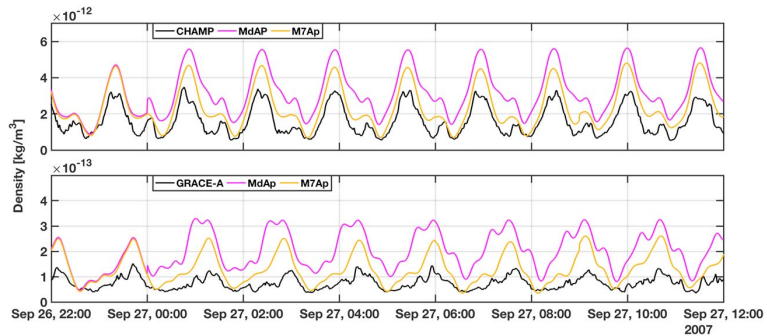

**Figure B2.** NRLMSISE-00 model estimates compared to CHAMP and GRACE-A accelerometer-derived densities. MdAp run is driven with daily average of 3-hr $Ap$, and M7Ap is driven with the full required history of $Ap$. Average orbital heights of GRACE-A and CHAMP satellites during this period were 473 and 360 km, respectively. CHAMP = Challenging Minisatellite Payload. GRACE = Gravity Recovery and Climate Experiment.

**Acknowledgments**

This research was partially supported by a research scholarship awarded to T. Kodikara by the Cooperative Research Centre for Space Environment Management, SERC Limited, through the Australian Government's Cooperative Research Centre Programme. This research was also partially supported by an Australian Research Council Linkage grant (LP160100561) awarded to B. A. Carter and K. Zhang. This study was also partially supported by the National Natural Science Foundation of China (ID: 41730109), and the Jiangsu dual creative talents and Jiangsu dual creative teams programme projects awarded in 2017. This research was undertaken with the assistance of resources from the National Computational Infrastructure (NCI), which is supported by the Australian Government. The European Space Agency is acknowledged for providing the Swarm-C data (product: SW_OPER_DNSCWND_2; version: L2PS-DEOS (5.83). The measurements plus $Kp$, $Ap$, and $F_{10.7}$ data are obtained from NOAA <www.ngdc.noaa.gov>. The IMF measurements are obtained from the OMNI database <omniweb.gsfc.nasa.gov>. SSI data stored at <lasp.colorado.edu/lisird/missions/timed> are the Level 3A postprocessed measurements from the SEE instrument on board the TIMED satellite. Eric Sutton is thanked for sharing CHAMP and GRACE accelerometer-derived densities. The National Center for Atmospheric Research (NCAR), Colorado, is acknowledged for making the TIE-GCM freely available at <www.hao.ucar.edu/modeling/tgcm/tie.php>. Sean Bruinsma is thanked for sharing the DTM-2013 model via <www.atmop.eu>. Mike Picone, Alan Hedin, and Doug Drob are thanked for sharing the NRLMSISE-00 model via <ccmc.gsfc.nasa.gov>. Stanley Solomon from NCAR is thanked for providing valuable advice at various stages of this study. Sean Elvidge is thanked for sharing the implementation for the modified Taylor diagram. Access to the model simulation data is provided via <goo.gl/R7Zehx>.

# References

Bennett, J., Sang, J., Smith, C., & Zhang, K. (2015). An analysis of very short-arc orbit determination for low-Earth objects using sparse optical and laser tracking data. *Advances in Space Research*, 55(2), 617–629. https://doi.org/10.1016/j.asr.2014.10.020

Bezděk, A. (2007). Lognormal distribution of the observed and modelled neutral thermospheric densities. *Studia Geophysica et Geodaetica*, 51(3), 461–468. https://doi.org/10.1007/s11200-007-0027-2

Bezděk, A., Sebera, J., & Klokočník, J. (2017). Validation of Swarm accelerometer data by modelled nongravitational forces. *Advances in Space Research*, 59, 2512–2521. https://doi.org/10.1016/j.asr.2017.02.037

Bruinsma, S. (2015). The DTM-2013 thermosphere model. *Journal of Space Weather and Space Climate*, 5(27), A1. https://doi.org/10.1051/swsc/2015001

Chen, J., Du, J., & Sang, J. (2017). Improved orbit prediction of LEO objects with calibrated atmospheric mass density model. *Journal of Spatial Science*, 0(0), 1–14. https://doi.org/10.1080/14498596.2017.1371089

Chen, Y., Liu, L., & Wan, W. (2011). Does the $F_{10.7}$ index correctly describe solar EUV flux during the deep solar minimum of 2007–2009? *Journal of Geophysical Research*, 116, A04304. https://doi.org/10.1029/2010JA016301

Codrescu, M. V., Fuller-Rowell, T. J., Munteanu, V., Minter, C. F., & Millward, G. H. (2008). Validation of the coupled thermosphere ionosphere plasmasphere electrodynamics model: CTIPE-mass spectrometer incoherent scatter temperature comparison. *Space Weather*, 6, S09005. https://doi.org/10.1029/2007SW000364

Dickinson, R. E., Ridley, E. C., & Roble, R. G. (1981). A three-dimensional general circulation model of the thermosphere. *Journal of Geophysical Research*, 86(A3), 1499–1512. https://doi.org/10.1029/JA086iA03p01499

Doornbos, E. (2012). *Thermospheric density and wind determination from satellite dynamics*. New York: Springer.

Drob, D. P., Emmert, J. T., Crowley, G., Picone, J. M., Shepherd, G. G., Skinner, W., et al. (2008). An empirical model of the Earth's horizontal wind fields: HWM07. *Journal of Geophysical Research*, 113, A12304. https://doi.org/10.1029/2008JA013668

Drob, D. P., Emmert, J. T., Meriwether, J. W., Makela, J. J., Doornbos, E., Conde, M., et al. (2015). An update to the Horizontal Wind Model (HWM): The quiet time thermosphere. *Earth and Space Science*, 2, 301–319. https://doi.org/10.1002/2014EA000089

Dudok de Wit, T., Bruinsma, S., & Shibasaki, K. (2014). On the use of modified Taylor diagrams to compare ionospheric assimilation models. *Journal of Space Weather and Space Climate*, 4, A06. https://doi.org/10.1051/swsc/2014003

Elvidge, S., Angling, M. J., & Nava, B. (2014). On the use of modified Taylor diagrams to compare ionospheric assimilation models. *Radio Science*, 49, 737–745. https://doi.org/10.1002/2014RS005435

Elvidge, S., Godinez, H. C., & Angling, M. J. (2016). Improved forecasting of thermospheric densities using multi-model ensembles. *Geoscientific Model Development*, 9(6), 2279–2292. https://doi.org/10.5194/gmd-9-2279-2016

Emmert, J. T. (2015). Thermospheric mass density: A review. *Advances in Space Research*, 56, 773–824. https://doi.org/10.1016/j.asr.2015.05.038

Emmert, J. T., McDonald, S. E., Drob, D. P., Meier, R. R., Lean, J. L., & Picone, J. M. (2014). Attribution of interminima changes in the global thermosphere and ionosphere. *Journal of Geophysical Research: Space Physics*, 119, 6657–6688. https://doi.org/10.1002/2013JA019484

Emmert, J. T., Warren, H. P., Segerman, A. M., Byers, J. M., & Picone, J. M. (2017). Propagation of atmospheric density errors to satellite orbits. *Advances in Space Research*, 59, 147–165. https://doi.org/10.1016/j.asr.2016.07.036

Friis-Christensen, E., Lühr, H., Knudsen, D., & Haagmans, R. (2008). Swarm—An Earth observation mission investigating geospace. *Advances in Space Research*, 41(1), 210–216. https://doi.org/10.1016/j.asr.2006.10.008

Fuller-Rowell, T. J., & Rees, D. (1980). A three-dimensional time-dependent global model of the thermosphere. *Journal of Atmospheric Sciences*, 37, 2545–2567. https://doi.org/10.1175/1520-0469(1980)037<2545:ATDTDG>2.0.CO;2

Hagan, M. E., Burrage, M. D., Forbes, J. M., Hackney, J., Randel, W. J., & Zhang, X. (1999). GSWM-98: Results for migrating solar tides. *Journal of Geophysical Research*, 104(A4), 6813–6827. https://doi.org/10.1029/1998JA900125

Hathaway, D. H. (2015). The solar cycle. *Living Reviews in Solar Physics*, 12, 4. https://doi.org/10.1007/lrsp-2015-4

Hedin, A. E. (1987). MSIS-86 thermospheric model. *Journal of Geophysical Research*, 92, 4649–4662. https://doi.org/10.1029/JA092iA05p04649

Hedin, A. E., Biondi, M. A., Burnside, R. G., Hernandez, G., Johnson, R. M., Killeen, T. L., et al. (1991). Revised global model of thermosphere winds using satellite and ground-based observations. *Journal of Geophysical Research*, 96(A5), 7657–7688. https://doi.org/10.1029/91JA00251

Heelis, R. A., Lowell, J. K., & Spiro, R. W. (1982). A model of the high-latitude ionospheric convection pattern. *Journal of Geophysical Research*, 87(A8), 6339–6345. https://doi.org/10.1029/JA087iA08p06339

Jones, M., Emmert, J. T., Drob, D. P., & Siskind, D. E. (2017). Middle atmosphere dynamical sources of the semiannual oscillation in the thermosphere and ionosphere. *Geophysical Research Letters*, 44, 12–21. https://doi.org/10.1002/2016GL071741

Knipp, D. J., Tobiska, W. K., & Emery, B. A. (2004). Direct and indirect thermospheric heating sources for solar cycles 21–23. *Solar Physics*, 224, 495–505. https://doi.org/10.1007/s11207-005-6393-4

Lin, C. S., Cable, S. B., Sutton, E. K., Marcos, F. A., Retterer, J. M., & Delay, S. H. (2013). Satellite drag validation of the thermosphere-ionosphere electrodynamics general circulation model (Tech. Rep. ADA587507). NM, USA: Air Force Research Laboratory.

Liu, H.-L. (2016). Variability and predictability of the space environment as related to lower atmosphere forcing. *Space Weather*, 14, 634–658. https://doi.org/10.1002/2016SW001450

Masutti, D., March, G., Ridley, A. J., & Thoemel, J. (2016). Effect of the solar activity variation on the global ionosphere thermosphere model (GITM). *Annales Geophysicae*, 34(9), 725–736. https://doi.org/10.5194/angeo-34-725-2016

Mayr, H. G., Harris, I., & Spencer, N. W. (1978). Some properties of upper atmosphere dynamics. *Reviews of Geophysics*, 16(4), 539–565. https://doi.org/10.1029/RG016i004p00539

McLaughlin, C. A., Krishna, D. M., Mehta, P. M., Lechtenberg, T., Hiatt, A., Fattig, E., & Locke, T. (2013). Thermosphere density variability, drag coefficients, and precision satellite orbits (Final Report). Defense Technical Information Center. AFRL-OSR-VA-TR-2013-0428. https://doi.org/10.21236/ada582025

McLaughlin, C. A., Manee, S., & Lechtenberg, T. F. (2011). Drag coefficient estimation in orbit determination. *The Journal of the Astronautical Sciences*, 58(3), 513–530. https://doi.org/10.1007/BF03321183

Menvielle, M., & Berthelier, A. (1991). The K-derived planetary indices: Description and availability. *Reviews of Geophysics*, 29(3), 415–432. https://doi.org/10.1029/91RG00994

Pedregosa, F., Varoquaux, G., Gramfort, A., Michel, V., Thirion, B., Grisel, O., et al. (2011). Scikit-learn: Machine learning in Python. *Journal of Machine Learning Research*, 12, 2825–2830.

Picone, J. M., Hedin, A. E., Drob, D. P., & Aikin, A. C. (2002). NRLMSISE-00 empirical model of the atmosphere: Statistical comparisons and scientific issues. *Journal of Geophysical Research*, 107(A12), 1468. https://doi.org/10.1029/2002JA009430

Qian, L., Burns, A. G., Emery, B. A., Foster, B., Lu, G., Maute, A., et al. (2014). *The NCAR TIE-GCM* (Chap. 7, pp. 73–83). Washington, DC: American Geophysical Union. https://doi.org/10.1002/9781118704417.ch7

Qian, L., & Solomon, S. C. (2012). Thermospheric density: An overview of temporal and spatial variations. *Space Science Reviews, 168*(1), 147–173. https://doi.org/10.1007/s11214-011-9810-z

Qian, L., Solomon, S. C., & Kane, T. J. (2009). Seasonal variation of thermospheric density and composition. *Journal of Geophysical Research, 114*, A01312. https://doi.org/10.1029/2008JA013643

Reber, C. A., & Hays, P. B. (1973). Thermospheric wind effects on the distribution of helium and argon in the Earth's upper atmosphere. *Journal of Geophysical Research, 78*(16), 2977–2991. https://doi.org/10.1029/JA078i016p02977

Richards, P. G., Fennelly, J. A., & Torr, D. G. (1994). EUVAC: A solar EUV flux model for aeronomic calculations. *Journal of Geophysical Research, 99*, 8981–8992. https://doi.org/10.1029/94JA00518

Richmond, A. D., Ridley, E. C., & Roble, R. G. (1992). A thermosphere/ionosphere general circulation model with coupled electrodynamics. *Geophysical Research Letters, 19*, 601–604. https://doi.org/10.1029/92GL00401

Ridley, A. J., Deng, Y., & Tóth, G. (2006). The global ionosphere thermosphere model. *Journal of Atmospheric and Solar-Terrestrial Physics, 68*, 839–864. https://doi.org/10.1016/j.jastp.2006.01.008

Siemes, C., de Teixeira da Encarnação, J., Doornbos, E., van den IJssel, J., Kraus, J., & Peréstý, R. (2016). Swarm accelerometer data processing from raw accelerations to thermospheric neutral densities. *Earth, Planets and Space, 68*(1), 1–16. https://doi.org/10.1186/s40623-016-0474-5

Siskind, D. E., Drob, D. P., Dymond, K. F., & McCormack, J. P. (2014). Simulations of the effects of vertical transport on the thermosphere and ionosphere using two coupled models. *Journal of Geophysical Research: Space Physics, 119*, 1172–1185. https://doi.org/10.1002/2013JA019116

Solomon, S. C., & Qian, L. (2005). Solar extreme-ultraviolet irradiance for general circulation models. *Journal of Geophysical Research, 110*, A10306. https://doi.org/10.1029/2005JA011160

Solomon, S. C., Qian, L., Didkovsky, L. V., Viereck, R. A., & Woods, T. N. (2011). Causes of low thermospheric density during the 2007–2009 solar minimum. *Journal of Geophysical Research, 116*, A00H07. https://doi.org/10.1029/2011JA016508

Sutton, E. (2009). Normalized force coefficients for satellites with elongated shapes. *Journal of Spacecraft and Rockets, 46*(1), 112–116. https://doi.org/10.2514/1.40940

Sutton, E. K. (2016). Interhemispheric transport of light neutral species in the thermosphere. *Geophysical Research Letters, 43*, 12,325–12,332. https://doi.org/10.1002/2016GL071679

Sutton, E. K., Thayer, J. P., Wang, W., Solomon, S. C., Liu, X., & Foster, B. T. (2015). A self-consistent model of helium in the thermosphere. *Journal of Geophysical Research: Space Physics, 120*, 6884–6900. https://doi.org/10.1002/2015JA021223

Tapping, K. F. (2013). The 10.7 cm solar radio flux ($F_{10.7}$). *Space Weather, 11*, 394–406. https://doi.org/10.1002/swe.20064

Taylor, K. E. (2001). Summarizing multiple aspects of model performance in a single diagram. *Journal of Geophysical Research, 106*(D7), 7183–7192. https://doi.org/10.1029/2000JD900719

Vallado, D. A. (2004). *Fundamentals of astrodynamics and applications* (Vol. 12). California: Space Technology Library.

Vallado, D. A., & Finkleman, D. (2014). A critical assessment of satellite drag and atmospheric density modeling. *Acta Astronautica, 95*, 141–165.

Walker, J. C. G. (1965). Analytic representation of upper atmosphere densities based on Jacchia's static diffusion models. *Journal of Atmospheric Sciences, 22*, 462–462. https://doi.org/10.1175/1520-0469(1965)022<0462:AROUAD>2.0.CO;2

Weimer, D. R. (2005). Predicting surface geomagnetic variations using ionospheric electrodynamic models. *Journal of Geophysical Research, 110*, A12307. https://doi.org/10.1029/2005JA011270

Wu, Q., Emery, B. A., Shepherd, S. G., Ruohoniemi, J. M., Frissell, N. A., & Semeter, J. (2015). High-latitude thermospheric wind observations and simulations with SuperDARN data driven NCAR TIEGCM during the December 2006 magnetic storm. *Journal of Geophysical Research: Space Physics, 120*, 6021–6028. https://doi.org/10.1002/2015JA021026

Xiong, C., Stolle, C., & Lühr, H. (2016). The Swarm satellite loss of GPS signal and its relation to ionospheric plasma irregularities. *Space Weather, 14*, 563–577. https://doi.org/10.1002/2016SW001439

Zesta, E., & Huang, C. (2016). *Satellite orbital drag* (pp. 329–347). Boca Raton, FL: CRC Press. https://doi.org/10.1201/9781315368474-19